\documentclass[]{aastex631}

\usepackage[T1]{fontenc}        

\received{2022 December 16}
\accepted{2023 February 26}

\submitjournal{ApJ}

\shorttitle{Varying Calcium Abundance}
\shortauthors{Sylwester et al.}

\graphicspath{{./}{figures/}}

\begin{document}

\title{Varying Calcium Abundances in Solar Flares seen by Solar Maximum Mission}


\correspondingauthor{K. J. H. Phillips}
\email{kennethjhphillips@yahoo.com}

\author[0000-0001-8428-4626]{B. Sylwester}
\affiliation{Space Research Centre, Polish Academy of Sciences (CBK PAN), Warsaw, Bartycka 18A, Poland}
\email{bs@cbk.pan.wroc.pl}

\author[0000-0002-8060-0043]{J. Sylwester}
\affiliation{Space Research Centre, Polish Academy of Sciences (CBK PAN), Warsaw, Bartycka 18A, Poland}
\email{js@cbk.pan.wroc.pl}

\author[0000-0002-3790-990X]{K. J. H. Phillips}
\affiliation{Scientific Associate, Earth Sciences Department, Natural History Museum, Cromwell Road, London SW7 5BD, UK}
\email{kennethjhphillips@yahoo.com}

\author[0000-0002-5299-5404]{A. K\k{e}pa}
\affiliation{Space Research Centre, Polish Academy of Sciences (CBK PAN), Warsaw, Bartycka 18A, Poland}
\email{ak@cbk.pan.wroc.pl}

\begin{abstract}

We report on calcium abundance $A({\rm Ca})$ estimates during the decay phases of 194 solar X-ray flares using archived data from the Bent Crystal Spectrometer (BCS) on Solar Maximum Mission (operational 1980~--~1989). The abundances are derived from the ratio of the total calcium X-ray line emission in BCS channel~1 to that in neighboring continuum, with temperature from a satellite-to-resonance line ratio. Generally the calcium abundance is found to be about three times the photospheric abundance, as previously found, indicating a ``FIP'' (first ionization potential) effect for calcium which has a relatively low FIP value. The precision of the abundance estimates (referred to hydrogen on a logarithmic scale with $A({\rm H}) = 12$), is typically $\sim \pm 0.01$, enabling any time variations of $A({\rm Ca})$ during the flare decay to be examined. For a total of 270 short time segments with $A({\rm Ca})$ determined to better than 2.3\% accuracy, many (106; 39\%) showed variations in $A({\rm Ca})$ at the $3\sigma$ level.  For the majority, 74 (70\%) of these 106 segments $A({\rm Ca})$ decreased with time, and  for 32 (30\%) $A({\rm Ca})$ increased with time. For 79 out of 270 (29\%) we observed constant or nearly constant $A({\rm Ca})$, and the remaining 85 (31\%) with irregular time behavior. A common feature was the presence of discontinuities in the time behavior of $A({\rm Ca})$. Relating these results to the ponderomotive force theory of Laming, we attribute the nature of varying $A({\rm Ca})$ to the emergence of loop structures in addition to the initial main loop, each with its characteristic calcium abundance.
\end{abstract}


\section{Introduction}\label{sec:intro}

In a recent series of papers \citep{rap17,jsyl20,jsyl22}, we investigated X-ray spectra of He-like calcium (\ion{Ca}{19}) near 3.17~\AA\ emitted by hot solar flare plasmas and observed by  the Bent Crystal Spectrometer (BCS) on the NASA Solar Maximum Mission (SMM: operational 1980 and 1984~--~1989). Solar Maximum Mission was launched in 1980 and operated throughout most of 1980 and again, after repair of attitude control units, from 1984--1989. The spacecraft and its complement of instruments including the X-ray Polychromator that BCS was part of was described by \cite{boh80}. The BCS was originally described by \cite{act80}, with revisions to instrument parameters given in recent studies by \cite{rap17} and \cite{jsyl20}. The BCS had eight channels, each with a curved crystal diffracting incoming X-rays from solar flares, with the diffracted radiation detected by position-sensitive proportional counters. For BCS channel~1, the wavelength range was nominally 3.165~--~3.231~\AA, and included the He-like Ca (\ion{Ca}{19}) resonance line and neighboring dielectronic satellites as well as continuum emission. A grid collimator in front of the crystals ensured that only flares from individual active regions were detected, although the main flare emission was not necessarily along the collimator central axis or boresight. Complete spectra over time intervals of a few seconds were obtained in the normal BCS operating mode, with higher time resolution on some occasions allowing spectra from very rapidly developing flares to be observed. The spectral resolution for this channel was 0.764~m\AA\ in 1980, and although this degraded by about 25\% in later stages of the SMM lifetime, the BCS was the highest-resolution spectrometer in this range of any spectrometer, solar or non-solar, since. We have availed ourselves of an archive of BCS and other SMM data that has been maintained on a NASA ftp site, and which remains an important source of information for studying flares.

Of particular interest has been the abundance of calcium in flare plasmas. This follows earlier analyses of BCS spectra \citep{jsyl84,jsyl98} in which strong evidence for flare-to-flare variations was presented. The recent re-assessment of instrument parameters by \cite{jsyl20} included the observation of deformities in the crystal surface leading to significant wavelength-dependent corrections in the crystal reflectivities. The corrected spectra showed a much improved agreement with the theoretical spectra for given temperatures, as calculated by \cite{phi18} with a purpose-written program having excitation data for the principal He-like calcium lines and associated dielectronic satellite lines. The theoretical spectra include ionized argon lines which make a minor contribution. A noticeable improvement in the agreement of observed and calculated spectra is in the ratio of the two \ion{Ca}{19} intercombination lines $x$ and $y$ (for list of lines, atomic transitions, and notation, see Table~1 of \cite{phi18}), observed to be 1:1.3 without the corrections (as discussed by \cite{phirainnie04}) but with the corrections is approximately 1:1, as observed.

\cite{jsyl22} analyzed BCS spectra during the decays of 194 flares seen by the BCS over the period of SMM operations, taking the ratios of the total emission in all lines included in channel~1 to the continuum which are a measure of the calcium abundance. This analysis took account of the revised instrument data as well as other improvements including the evaluation of a slow change of the wavelength dispersion over the spacecraft lifetime. Taking only the decay phases of flares avoids the complications of multiple sources with different temperatures in the often complex rise phases. An isothermal assumption for the flare decay emission was found to give a good approximation to the BCS spectral data which cover only a small wavelength interval. Temperatures for each BCS spectrum can be found from satellite-to-resonance line intensity ratios but also from the emission ratio of the two channels (0.5~--~4~\AA, 1~--~8~\AA) of the Geostationary Operational Environmental Satellites (GOES) with pre-flare background subtracted, the two temperatures being found to be closely related. For most of the 194 flare decays, subintervals could be defined, typically a few minutes long, in which lines-to-continuum ratios and therefore calcium abundances estimated. The derived distribution of calcium abundances $A({\rm Ca})$ (expressed logarithmically with $A({\rm H}) = 12$) for the total of 2806 subintervals was close to a Gaussian, with maximum 6.74 and standard deviation 0.09. As this averaged abundance is about a factor 3 higher than photospheric or meteoritic calcium abundances (approximately $A({\rm Ca}) = 6.3$: see \cite{king20,asp21,lod21}), a ``FIP'' (first ionization potential) effect (an enhancement of flare or coronal abundances over photospheric by about a factor 3 for elements with FIP $\lesssim 10$~eV) was thus indicated for Ca (FIP $= 6.1$~eV).

Individual estimates of calcium abundances $A({\rm Ca})$ over short time intervals during their decays made by \cite{jsyl22} have typical uncertainties as small as $\pm 0.01$, particularly for medium-to-large flares (GOES X-ray importance M or X). This precision allows changes in $A({\rm Ca})$ with time during flare decays to be examined and as a result time trends -- increasing, decreasing, both increasing and decreasing, or approximately constant -- were identified. The changes in the values of $A({\rm Ca})$ were found to be as high as 0.02, and so evidently the width of the Gaussian distribution (standard deviation 0.09) found by \cite{jsyl22} was due to time variations rather than statistical uncertainties.

The analysis methods for this work are described below in Section~\ref{sec:BCS_Data_Anal} and the resulting abundance measurements are given in Section~\ref{sec:Calcium_Abund}. A large proportion of flare decays showed significantly varying values of $A({\rm Ca})$, and it was found that sudden changes in $A({\rm Ca})$ occurred simultaneously with identifiable features in temperatures and positional information (described in Section~\ref{sec:GOES_BCScontm}) which though small appear to be due to new emitting structures as each flare evolves. A detailed discussion of some example flares in Section~\ref{sec:Example_Flares} and a summary of the characteristics of all flares (Section~\ref{sec:Abund_Meas}) are given. In Section~\ref{sec:Discussion} we compare our results to other solar flare abundance studies and also show how our results relate to theories attempting to explain the FIP effect, in particular that of the ponderomotive force idea of \cite{lam21}. We summarize this study in Section~\ref{sec:Summary}.

\section{BCS Data Analysis} \label{sec:BCS_Data_Anal}

\subsection{Procedures}\label{sec:Procedures}
Data from the Bent Crystal Spectrometer in the SMM archive consist of series of spectra in the form of photon counts in 254 wavelength bins, the spectra being taken over data gathering time intervals from 9 to 20~seconds depending on the X-ray emission level and a pre-determined operation mode. The grid collimator through which X-rays were incident had a square field of view with a full width half maximum (FWHM) of 6~arcminutes, as was measured pre-launch and subsequently deduced from in-flight observations \citep{jsyl20}. Of the eight channels, the channel~1 wavelength range (3.165~--~3.231~\AA\ for a source along the BCS boresight) included the He-like Ca (\ion{Ca}{19}) resonance line $w$ and other \ion{Ca}{19} associated lines and \ion{Ca}{18} satellite lines emitted during flares. The emission ratios of dielectronically formed \ion{Ca}{18} satellites, in particular satellite $k$, to \ion{Ca}{19} line $w$ are inversely related to temperature (for notation and other details see \cite{gab72} and Table~1 of \cite{phi18}); we denote the temperature from this ratio by $T_{k/w}$. A background radiation in channel~1 spectra is due to flare free--free and free--bound continua, with negligible amounts due to instrumental causes, in particular fluorescence of the crystal material by solar X-rays. The high solar activity of solar cycle 21 (peaking in December 1979 shortly before SMM launch) and cycle 22 (peaking in November 1989) meant that over much of the SMM period several active regions were usually present on any given day. A selection was made daily during the operations period for the spacecraft to be pointed at active regions predicted to be most flare-prolific.

Here we use the same set of flare spectra that was described by \cite{jsyl22}, consisting of 2806 averaged spectra in subintervals during the decay phases of 194 flares. The maxima of the flare decays have a wide range of GOES X-ray importance: 2 of B class; 87 of C class; 91 of M class; and 14 of X class. The time spans of the decay phases were selected from an inspection of BCS and GOES (1~--~8~\AA\ channel) X-ray light curves, avoiding any instrumental anomalies in either GOES or BCS. The light curves were not always simple, and quite often further increases in X-ray emission occurred soon after a first maximum. For a very small number of flares, the GOES light curve reached the GOES instrument threshold, but these periods did not overlap with the BCS spectral observations used here. As described by \cite{jsyl22}, the BCS spectra were corrected for detector saturation. Details of flares analysed and their characteristics are available from a web site ({\tt http://www.cbk.pan.wroc.pl/js/ApJ\_930\_77\_2022/Table\_1.txt}); the first ten lines of the list were given as Table~1 of \cite{jsyl22}.

\subsection{GOES Data Analysis}\label{sec:GOES_Analysis}
The flare decays analyzed cover a nine-year period during which GOES data were available from successively GOES-2, GOES-5, and GOES-6. A number of problems were identified in the analysis of these data, including sudden discontinuous jumps in the light curves, unrelated between the two (0.5~--~4~\AA\ and 1~--~8~\AA) channels, or short sections of light curves shifted up or down by constant amounts. Corrections to these were done by hand wherever possible, ensuring that the corrections did not introduce discontinuities in temperatures and emission measures derived from the emission ratio of the two channels. After these corrections were made, the irradiances (units of W~m$^{-2}$) in the two GOES channels were smoothed using a 60-second box-car filter in order to remove any high-frequency noise with non-solar or instrumental origin.

Since GOES views X-rays from the entire Sun, there was generally some contribution to the flare X-ray emission from elsewhere on the Sun outside the BCS collimator field of view. Thus, for comparison with BCS channel~1 emission, it was necessary for a background level from either before or after (or an average of both) to be subtracted for both GOES channels. This background level could be determined in a straightforward way for flares with simple light curves (no significant emission from other flares), but for two or more flares with overlapping light curves, the pre-flare or post-flare emission levels (or both) were more difficult to define. As indicated by \cite{jsyl22}, temperatures from the emission ratio of the two GOES channels were derived by taking the background-subtracted emission in both GOES channels for all subintervals; this is here denoted by $T_{\rm Gb}$ and the corresponding emission measure $EM_{\rm Gb}$ (equal to $N_e^2 V$  where $N_e$ is the electron density and $V$ the emitting volume). Our earlier analysis (Figure~7 of \cite{jsyl22}) showed a very close relation of $T_{\rm Gb}$ with the BCS temperature $T_{k/w}$ for flares with simple light curves, but a less clear relation for flares with more complex light curves. In either case, values of $T_{k/w}$ were found to be $1 - 2$~MK higher than $T_{\rm Gb}$.

Over the SMM operational period, BCS spectra were available for analysis for nearly all times apart from when the spacecraft fine pointing was lost (from 1980 November to 1984 April). Use of GOES irradiance data was needed for our analysis to enable comparison with spectra from BCS channel~1, but as mentioned over the nine-year time span GOES irradiances were measured from three different spacecraft (GOES-2, GOES-5 and GOES-6). As was evident from overlapping time periods, the detectors on board the three spacecraft had slightly different sensitivities. We used the following relations to bring the observed emission in each of the three satellites into a unified system. Empirical relations for an irradiance $f_{\rm G}$ were defined in such a way that combines the irradiance in the 0.5~--~4~\AA\ channel ($f_{\rm G4}$) and that in the 1~--~\AA\ channel ($f_{\rm G8}$):

\begin{equation}
{\rm log}\,\,f_{\rm G} = {\rm log} (f_{{\rm G4}} + 0.07 \times f_{{\rm G8}}) - 0.2,
\label{eq:logf_2}
\end{equation}
\noindent for GOES-2,

\begin{equation}
{\rm log}\,\,f_{\rm G} = {\rm log} (f_{{\rm G4}} + 0.06 \times f_{{\rm G8}}) - 0.054,
\label{eq:logf_5}
\end{equation}

\noindent for GOES-5 operational in 1984, and

\begin{equation}
{\rm log}\,\,f_{\rm G}= {\rm log} (f_{{\rm G4}} + 0.065 \times f_{{\rm G8}}) - 0.05,
\label{eq:logf_6}
\end{equation}

\noindent for GOES-6 operational from 1984 to the end of SMM operations in 1989. In each of these relations, GOES irradiances are expressed in W~m$^{-2}$.

\subsection{GOES and BCS Channel 1 Emission}\label{sec:GOES_BCScontm}
\begin{figure}
\centerline{\includegraphics[width=0.99\textwidth,clip=]{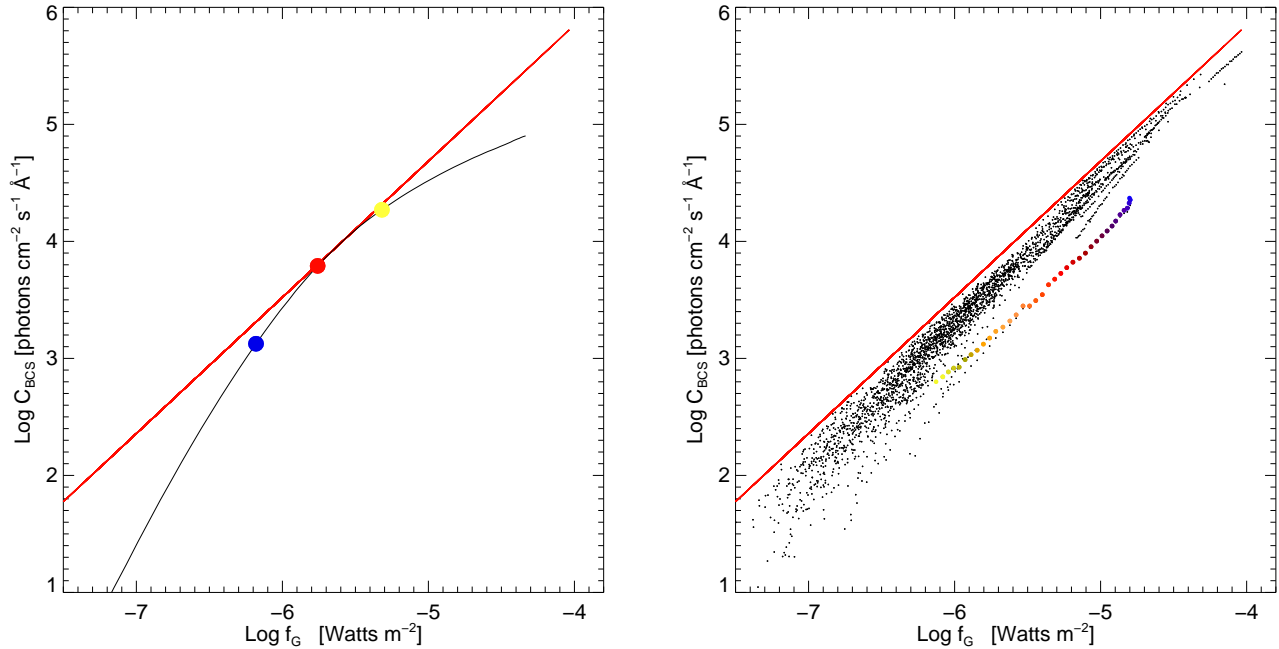}}
\caption{Left panel: Logarithmic plot of the BCS continuum irradiance at $\lambda=3.1723$~\AA\ against GOES irradiance. The curved black line is the dependence of $C_{\rm BCS}$ on $f_{\rm G}$ from calculated spectra \citep{phi18} for constant emission measure ($EM=10^{48}$~cm$^{-3}$) and changing temperature in the range 3--100~MK. The blue, red and yellow points represent values for temperatures of 7, 10, and 15~MK respectively. For $T = 10$~MK the curve meets the red line. For sources at $T{\simeq}10$~MK and varying $EM$, the red point would move along the red line. Right panel: {BCS observations for subintervals during the 188 flares analyzed here for which BCS and GOES data are both available given as a a scatter plot of $C_{BCS}$ against $f_G$ for spectra of all 2698 subintervals.} Practically all the points are at least slightly displaced from the BCS boresight, as indicated by the fact that they do not perfectly align with the solid red line. The colored dots are for flare~53, for which there is a relatively large displacement of about 0.75 (see text).
\label{fig:fig1}}
\end{figure}

The GOES irradiance is related to the emission in BCS channel~1 as has been illustrated by the similarity of the temperatures $T_{k/w}$ and $T_{\rm Gb}$ (Figure~7 of \cite{jsyl22}). With the GOES irradiance $f_{\rm G}$ for any time during the SMM operational period, it is possible to estimate the BCS channel~1 continuum. Figure~\ref{fig:fig1} illustrates the method by which this is done. In the left panel, the BCS channel~1 continuum irradiance $C_{\rm BCS}$ at $\lambda=3.1723$~\AA\ (on the short-wavelength side of \ion{Ca}{19} line $w$) is plotted logarithmically against $f_{\rm G}$ (abscissa) for a constant emission measure equal to $10^{48}$~cm$^{-3}$ and temperatures varying over the range 3~--~100~MK (curved black line). The values of $C_{\rm BCS}$ were obtained from the calculated spectra of \cite{phi18}. The red line is identical to that in the right panel, and is empirically determined to be $C_{\rm BCS} = 1.16 \times f_{\rm G} + 10.5$. The large colored dots are for temperatures of 7~MK (blue), 10~MK (red), and 15~MK (yellow) respectively. For a temperature of 10~MK and varying emission measure, the red point would move along the red diagonal line. The right panel shows the same red diagonal line but with data points added for all spectra analyzed here. For each flare decay, the points move diagonally downward from maximum values of $C_{\rm BCS}$ and $f_{\rm G}$. There is an apparent convergence of the points from an upward direction toward the red line (one point actually exceeds the line but is still within error bars). The small colored dots trace out the path for flare no.~53 of our analysis (SOL1980-Nov-13T01:03~UT), coded blue for earlier times, red and yellow for later times.  As can be seen, the points progress downward parallel to the red line but displaced from it by an amount equal to $\sim 0.75$. We attribute this displacement to the offset of the flare from the BCS boresight. Flare no.~53 had an exceptionally large offset (approximately 4~arcminutes) from the BCS boresight; the remaining points in Figure~\ref{fig:fig1} had offsets of 2~arcminutes or less.

\begin{figure}
\vspace{3mm}
\centerline{\includegraphics[width=1.0\textwidth,clip=]{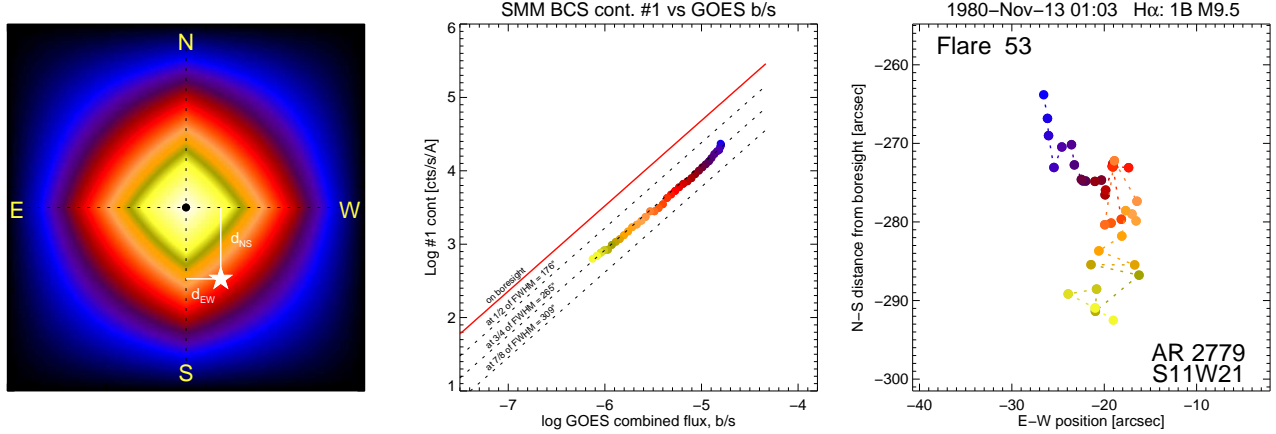}}
\caption{The determination of the displacement of flare emission along the  N~--~S direction. Left panel: The BCS collimator pattern projected on to the sky with orientation shown and an X-ray source at an arbitrary position shown as a white asterix (bottom right or South and West of the BCS boresight). The E~--~W position  $d_{\rm EW}$ is defined by the displacement of the BCS spectrum from its position relative to a flare along the BCS boresight in the dispersion plane; the N~--~S position $d_{\rm NS}$ is obtained from the {\it vertical distance} from the red line in the middle panel of this figure showing (as colored, time coded points) the logarithm of BCS channel~1 continuum emission observed for the flare No.~53 (SOL1980-Nov-13T01:03~UT) plotted against the logarithm of combined GOES flux (Equation~\ref{eq:logf_2}), with the red line showing the dependence for a flare located exactly along the BCS boresight. The vignetting factor 0.75 is that by which the BCS continuum emission is lower than the value for a flare along the BCS boresight. Right panel: The position of flare no.~53 deduced from the spectral shifts (E~--~W direction) and reduction in total emission deduced from comparison with GOES emission (N~--~S direction) within the BCS collimator field of view (time-coded colored dots). The magnitude of the N~--~S direction is determined but not the sign (N or S).
\label{fig:fig2}}
\end{figure}

Figure~\ref{fig:fig2} illustrates where flare no.~53 is located (obtained from H$\alpha$ heliographic coordinates) in the left-most panel with $d_{\rm EW}$ and $d_{\rm NS}$ offsets in the E~--~W and N~--~S directions respectively indicated. The middle panel shows the colored points given in Figure~\ref{fig:fig1} but with the vignetting factors arising from the displacement of the flare from the BCS boresight, in this case 0.75, which is known from the flare's heliographic coordinates. If the coordinates were unknown, comparison of the total BCS emission with that given by the temperature and emission measure would have enabled the flare's position to be known: first $d_{\rm EW}$ (the E~--~W position) is readily available from the displacement of the channel~1 spectrum from an on-axis flare, then $d_{\rm NS}$ (the N~--~S position) from the comparison of the BCS continuum emission (specifically $C_{\rm BCS}$) and the GOES irradiance $f_{\rm G}$ as defined by Equations~\ref{eq:logf_2} -- \ref{eq:logf_6}. Note, however, that the sign of the flare's N~--~S position remains ambiguous since this is determined only from the absolute value of the BCS source offset from the collimator boresight. In the right-most panel of Figure~\ref{fig:fig2} the deduced positions of flare no.~53 are given in the familiar solar coordinates ($x$ or E~--~W position with W positive, $y$ or N~--~S position  with N positive).

For flares late in the SMM operations period, the BCS continuum emission at 3.1723~\AA\ could not be determined as the short-wavelength region was outside the channel~1 spectral range. In this case, continuum emission on the short-wavelength side of \ion{Ca}{19} line~$z$, 3.2157~\AA, was used instead.

%

%

Apart from obtaining positional information, relating BCS continuum emission to GOES emission allows us to investigate possible detector saturation effects. The period of SMM operations coincided with some very high solar activity with flares having GOES classes of X1 or greater. As the BCS was known to be subject to saturation effects during some exceptionally powerful flares in 1984 April, we examined whether saturation effects for other flares also that might affect the present analysis. There is a small  departure from non-linearity at low and high values of either GOES or channel~1 continuum emission which can be attributed to saturation effects. An appropriate correction for saturation was applied to all BCS spectra as a first step in the data reduction.

\section{Calcium Abundance Results}\label{sec:Calcium_Abund}

\subsection{Abundance Measurements}\label{sec:Abund_Meas}

Using corrected GOES irradiances and deriving background-subtracted temperatures $T_{\rm Gb}$ and emission measures $EM_{\rm Gb}$, we constructed for each flare decay a so-called diagnostic diagram (DD), i.e. a plot of log~$T_{\rm Gb}$ against $(1/2)\, {\rm log} \, EM_{\rm Gb}$, the latter being a rough measure of $N_e$ on the assumption of constant $V$. This was used in a previous work \citep{jak92} for studying the flare thermodynamic evolution with model flares from the Palermo--Harvard hydrodynamic code. As will be shown, this helps greatly in identifying sections of the light curves of some flares that show discontinuities in derived calcium abundance determinations which we identify with emergence of different structures as each flare decays.

Time trends of $A({\rm Ca})$ with uncertainties were calculated during the 194 flare decays over periods, called here time segments. Of a total of 302 time segments, some 270 time segments in 159 flare decays were identified in which the trends could be found to within 2.3\%, the remainder having too few time points or too uncertain values of derived abundances. In Table~\ref{tab:notable_trends} we list characteristics of a selection of these segments. The first ten rows give data for flare decays that were particularly well observed. They are discussed in Section~\ref{sec:Example_Flares} and are illustrated in Figures~\ref{fig:fig3}--\ref{fig:fig6}. The following rows in the table give data for five flare decays in each of three categories: those with the least changing values of $A({\rm Ca})$, those with the largest values of increasing $A({\rm Ca})$, and those with the largest values of decreasing $A({\rm Ca})$. Dates, times (UT), NOAA AR No. and location, GOES class, and derived calcium abundances $A({\rm Ca})$ with time derivatives ($d[A({\rm Ca)]}/dt$, units of minute$^{-1}$, are given for each flare. These flares are discussed in Section~\ref{sec:Trends_for_all_flares}. A full version of Table~\ref{tab:notable_trends} including all flare decays analyzed is available on-line at {\tt http://www.cbk.pan.wroc.pl/js/ApJ\_2022/Table\_Segments.txt}

\begin{deluxetable*}{ccccccccccrc}
\tabletypesize{\small}
\tablecaption{Time Segments in BCS flare decays with Calcium Abundances and Time Trends \label{tab:notable_trends} }
\tablewidth{0pt}
\tablehead{\colhead{No.}  & \colhead{Date and Time (UT)} & \colhead{NOAA Active} & \colhead{AR Location} &  \colhead{GOES} & \colhead{nS} & \colhead{Segment Time} & \colhead{$A({\rm Ca})$} & \colhead{Time trend of}  \\
&\colhead{of X-ray maximum}&\colhead{Region}&\colhead{(Heliogr. Coords.)}&\colhead{Class} & &\colhead{(UT)} & \colhead{with uncert.} & \colhead{$A({\rm Ca})$ with uncert.} \\
&&&&&&&&(minute$^{-1}$)\\}
\startdata\\
\multicolumn{8}{l}{\it Illustrated in Figures~\ref{fig:fig3} to \ref{fig:fig6} }\\                    \\
10&1980-May-21~~~21:07& 2456 & S14W15 &  X1.5 & 16 & 21:15:38 & $6.68 \pm 0.01$ & $+0.0064 \pm 0.0035$ \\
10&1980-May-21~~~21:07& 2456 & S14W15 &  X1.5 &  7 & 21:26:26 & $6.71 \pm 0.01$ & $+0.0100 \pm 0.0138$ \\
 \\
 66&1984-May-19~~~21:53& 4492 & S10E66 &  X4.3 & 20 & 22:05:54 & $6.53 \pm 0.02$ & $-0.0325 \pm 0.0027$ \\
 66&1984-May-19~~~21:53& 4492 & S10E66 &  X4.3 & 16 & 22:19:44 & $6.58 \pm 0.04$ & $+0.0925 \pm 0.0065$ \\
 \\
159&1988-Feb-20~~~04:20& 4951 & S08W68 &  M1.3 & 20 & 04:26:38 & $6.84 \pm 0.03$ & $+0.0576 \pm 0.0053$ \\
159&1988-Feb-20~~~04:20& 4951 & S08W68 &  M1.3 &  8 & 04:39:32 & $6.84 \pm 0.01$ & $-0.0048 \pm 0.0134$ \\
159&1988-Feb-20~~~04:20& 4951 & S08W68 &  M1.3 & 13 & 04:52:59 & $6.83 \pm 0.03$ & $-0.0452 \pm 0.0107$ \\
\\
173&1988-Jun-24~~~05:27& 5047 & S18W45 &  M2.5 & 24 & 05:47:05 & $6.69 \pm 0.02$ & $-0.0205 \pm 0.0024$ \\
173&1988-Jun-24~~~05:27& 5047 & S18W45 &  M2.5 & 11 & 06:09:23 & $6.61 \pm 0.04$ & $-0.0574 \pm 0.0089$ \\
173&1988-Jun-24~~~05:27& 5047 & S18W45 &  M2.5 &  6 & 06:23:47 & $6.58 \pm 0.03$ & $-0.0663 \pm 0.0212$ \\
\\
\multicolumn{8}{l}{\it Flare decays with least changing $A({\rm Ca})$}                   \\
190&1989-Aug-16~~~01:15& 5629 & S18W84 & X13.0 & 30 & 04:57:28 & $6.70 \pm 0.01$ & $+0.00002\pm0.00006$ \\
 58&1984-May-02~~~19:27& 4474 & S11W58 &  M3.0 & 22 & 19:34:35 & $6.70 \pm 0.02$ & $-0.0001 \pm0.0004$ \\
 84&1985-Apr-24~~~09:36& 4647 & N06E27 &  X1.9 & 11 & 11:13:15 & $6.69 \pm 0.01$ & $-0.0001 \pm0.0007$ \\
157&1988-Jan-02~~~21:44& 4912 & S34W18 &  X1.5 & 23 & 00:53:04 & $6.69 \pm 0.03$ & $+0.0001 \pm0.0001$ \\
 83&1985-Apr-24~~~05:03& 4647 & N05E26 &  C7.3 & 14 & 05:09:09 & $6.72 \pm 0.01$ & $-0.0001 \pm0.0005$ \\
\\
\multicolumn{8}{l}{\it Flare decays with fastest increasing $A({\rm Ca})$}                   \\
117&1987-May-24~~~09:06& 4811 & N31W32 &  C1.5 &  3 & 09:15:09 & $6.68 \pm 0.05$ & $+0.0272 \pm0.0081$ \\
  8&1980-Apr-30~~~20:26& 2396 & S13W90 &  M2.2 &  3 & 20:37:47 & $6.74 \pm 0.04$ & $+0.0311 \pm0.0069$ \\
184&1989-Jun-04~~~02:20& 5517 & S19E34 &  M1.1 &  4 & 02:38:15 & $6.77 \pm 0.06$ & $+0.0327 \pm0.0078$ \\
161&1988-Mar-16~~~01:58& 4964 & S25E34 &  M1.1 &  4 & 02:04:20 & $6.78 \pm 0.08$ & $+0.0337 \pm0.0071$ \\
 28&1980-Aug-23~~~20:50& 2629 & N17W38 &  C5.2 &  6 & 20:54:44 & $6.82 \pm 0.09$ & $+0.0475 \pm0.0056$ \\
\\
\multicolumn{8}{l}{\it Flare decays with fastest decreasing $A({\rm Ca})$}      \\
142&1987-Aug-08~~~13:57& 4835 & S25W26 &  M1.4 &  3 & 14:09:48 & $6.73 \pm 0.04$ & $-0.0219 \pm0.0081$ \\
 77&1985-Jan-21~~~17:20& 4617 & S09W33 &  M1.4 &  3 & 17:27:32 & $6.62 \pm 0.02$ & $-0.0267 \pm0.0078$ \\
147&1987-Sep-05~~~00:21& 4849 & S25E32 &  C3.5 &  5 & 00:28:04 & $6.91 \pm 0.10$ & $-0.0284 \pm0.0080$ \\
101&1986-Oct-14~~~14:50& 4749 & S04W87 &  C2.2 &  4 & 15:04:29 & $6.71 \pm 0.08$ & $-0.0319 \pm0.0081$ \\
154&1987-Nov-26~~~03:20& 4891 & S21W48 &  C3.4 &  3 & 03:25:50 & $6.76 \pm 0.09$ & $-0.0492 \pm0.0088$ \\
\\
\enddata
\tablenotetext{}{Remarks:  No. (col. 1) gives the flare decay serial number in the on-line table (see the on-line web site {\tt http://www.cbk.pan.wroc.pl/js/ApJ\_930\_77\_2022/Table\_1.txt}).
\\nS (col. 7) is the number of subintervals in each flare segment.
\\Abundances (col. 9) $A({\rm Ca})$ are logarithmic with $A({\rm H})=12$. Stated uncertainties include any time variations. The trends in the last column are given for changes in $A({\rm Ca})$ per minute.}
\end{deluxetable*}

\subsection{Example flares}\label{sec:Example_Flares}

We illustrate the different time behavior of the $A({\rm Ca})$ variations found from this study by four example flare decays. For each of these, we show a four-panel figure consisting of (top left) the GOES 1~-~8~\AA\ irradiance (in W~m$^{-2}$, plotted logarithmically) and temperature $T_{\rm Gb}$ (temperature from the ratio of background-subtracted emission in the two GOES channels) against time (UT); (top right) the location of the dominant flaring source within the BCS field of view in arcseconds (E~--~W position from spectral shift, with west positive; N~--~S position from comparison of BCS and GOES emission, with north positive); (bottom left) diagnostic diagram as described in \cite{jak92}; and (bottom right) time variation of $A({\rm Ca})$ with uncertainties and linear fits to the trends. In each figure panel, the times of BCS observations in subintervals are indicated by colors, blue for the start of the selected observational sequence progressing via red and orange to yellow for the end of the sequence.

%
\begin{figure}
\centerline{\includegraphics[width=0.75\textwidth,clip=]{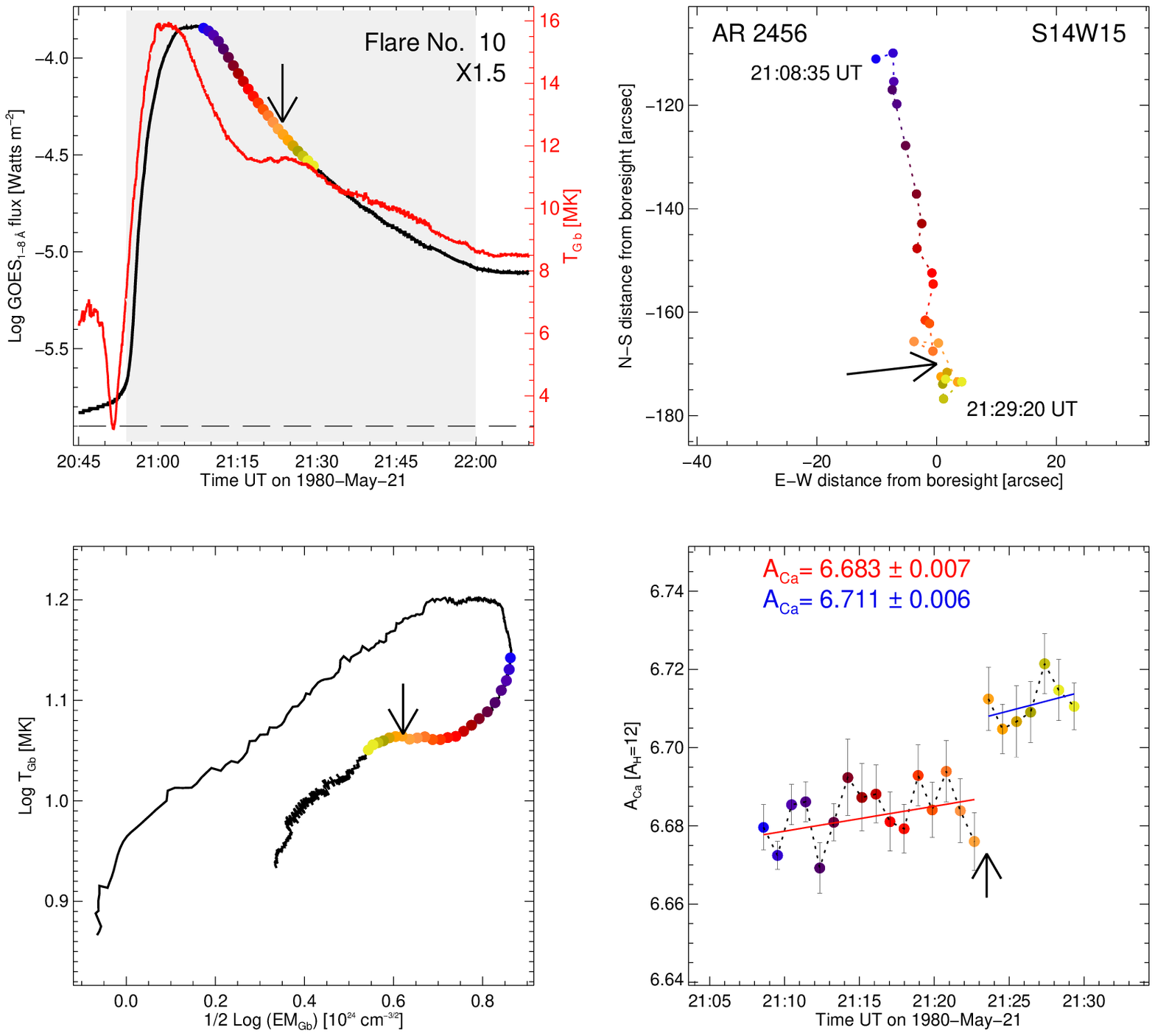}}
\caption{Results for the flare on 1980 May~21 (SOL1980-May-21T21:07~UT, no.~10 in the on-line Table).  Top left: GOES 1~--~8~\AA\ channel flux (black curve) and temperature $T_{\rm Gb}$ (red). The colored dots on the 1~-~8~\AA\ curve represent times of $A({\rm Ca})$ measured from BCS spectra (blue for earlier times, yellow and red later times). Top right: Flare locations in N~--~S direction (from comparison of BCS and GOES irradiances; the measurements have a north--south ambiguity) and E~--~W direction (west is positive: from displacement of BCS channel~1 spectra). Lower left: The diagnostic diagram or plot of $T_{\rm Gb}$ against $(1/2)\, \times {\rm log} (EM_{\rm Gb})$ from GOES irradiances, with colored dots showing the times of BCS observations.  Bottom right: Time plot of measured calcium abundances $A({\rm Ca})$ (error bars from statistical uncertainties in photon count rates). The red and blue lines are the best linear fits to the trends for individual time segments with an apparent break at 21:23:30~UT. Mean values of $A({\rm Ca})$ for these segments are given in colors corresponding to best-fit lines.
\label{fig:fig3}}
\end{figure}

One of the best observed flares in the SMM era was the X1.5 flare on 1980 May~21 (SOL1980-May-21T21:07~UT), peaking at about 21:07~UT from the disk active region AR2456, location S14W15. A review of observations of this flare was given by \cite{deJ85}. Results for this analysis are given in Figure~\ref{fig:fig3}. The position of the main emission seen in BCS channel~1 (top right panel) indicates a N~--~S shift of about 1~arcminute with little motion in the E~--~W direction.  The diagnostic diagram (bottom left panel) shows a discontinuity at around 21:23~UT. Although scarcely noticeable in the X-ray light curve, the temperature $T_{\rm Gb}$ indicates the appearance of a second emitting feature (top left panel). In the time plot of $A({\rm Ca})$ (bottom right panel) two segments lasting for 15 and 5 minutes can be distinguished, with sudden increase of 0.025 in $A({\rm Ca})$ at 21:23:30~UT; this amount is much larger than the $1\sigma$ error bars on individual points.  Table~\ref{tab:notable_trends} (first two rows) gives the abundance time trends and uncertainties for this flare. The irregularity in the diagnostic diagram occurs simultaneously with the $A({\rm Ca})$ increase.

Spatial information on the flares discussed here can be very roughly obtained from other instruments on SMM in particular the Flat Crystal Spectrometer (FCS), but  a time of several minutes was needed to complete scans during which most flares were likely to have significantly evolved in shape. For the case of the 1980 May~21 flare, images from the Hard X-ray Imaging Spectrometer (HXIS) are available during the flare decay. Over the period 20:53~UT to 23:05~UT there is a shift southward of the central part of the flare emission of 36~arcsecond. While there is qualitative agreement with the shift indicated by Figure~\ref{fig:fig3} (top right panel), the BCS shift is larger (65~arcsecond) over the shorter period of 21:08~--~21:38~UT. Agreement cannot therefore be claimed between the HXIS and BCS observations, although this might be at least partly explained by the low spatial resolution of the HXIS images (32~arcsecond for its maximum field of view: \cite{vanB80}).

%
\begin{figure}
\centerline{\includegraphics[width=0.75\textwidth,clip=]{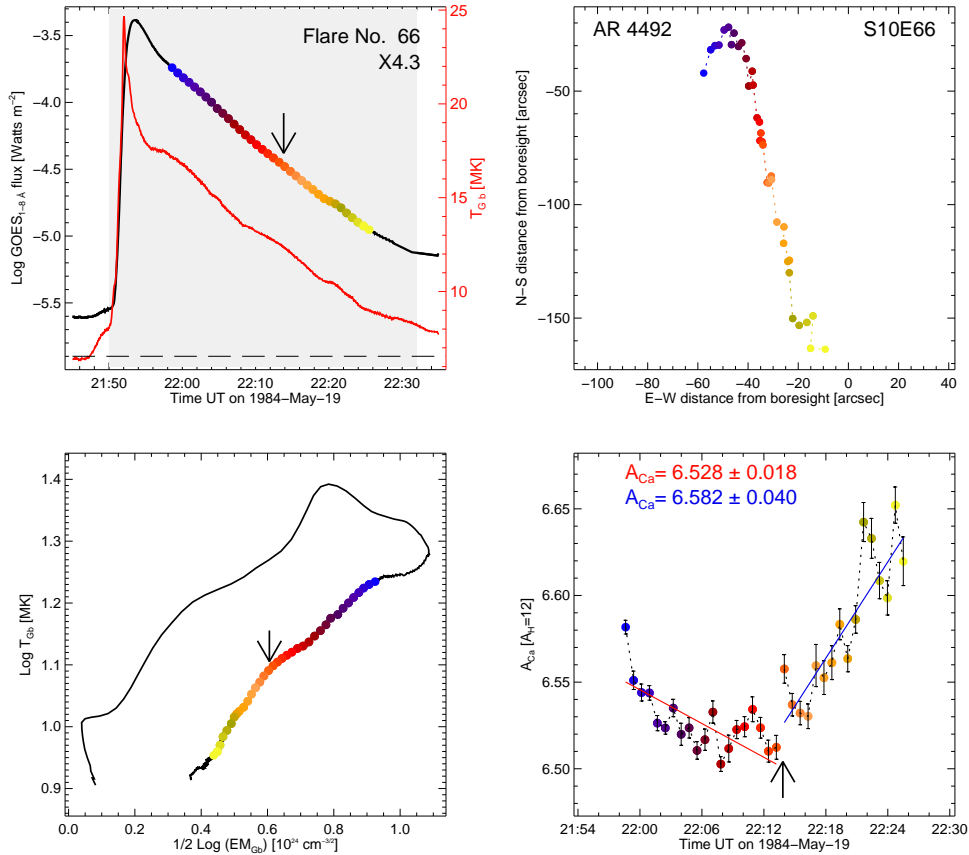}}
\caption{
Four-panel diagram (like Figure~\ref{fig:fig3}) for the near-limb flare on 1984 May~19 (SOL1984-May-19T21:53~UT: no.~66 in the on-line table and Table~\ref{tab:notable_trends}). The decrease followed by increase in measured $A({\rm Ca})$ values (bottom right panel) occur at a time 22:13:50~UT that is simultaneous with a direction change in the path of the points in the diagnostic diagram (arrowed in the bottom left panel).
\label{fig:fig4}}
\end{figure}

Figure~\ref{fig:fig4} shows the corresponding results for the X4.3 flare of 1984 May~19 (SOL1984-May-19T21:53~UT) near the south--east limb.  A motion approximately parallel to the limb of the main X-ray source over about 25~minutes is indicated by the BCS observations. As shown by the bottom right panel, $A({\rm Ca})$ significantly declined for about 12 minutes then rose toward the end of the observations. The mean values of $A({\rm Ca})$ are $6.527\pm 0.018$ for the first time segment and $6.582\pm 0.040$ for the second. The change from declining to rising $A({\rm Ca})$ occurs at about 22:10~UT, a time simultaneous with a fluctuation in the path indicated in the diagnostic diagram (marked by an arrow). As with the 1980 May~21 flare, the behavior is consistent with a second source that becomes dominant over the original source.

%
\begin{figure}
\centerline{\includegraphics[width=0.75\textwidth,clip=]{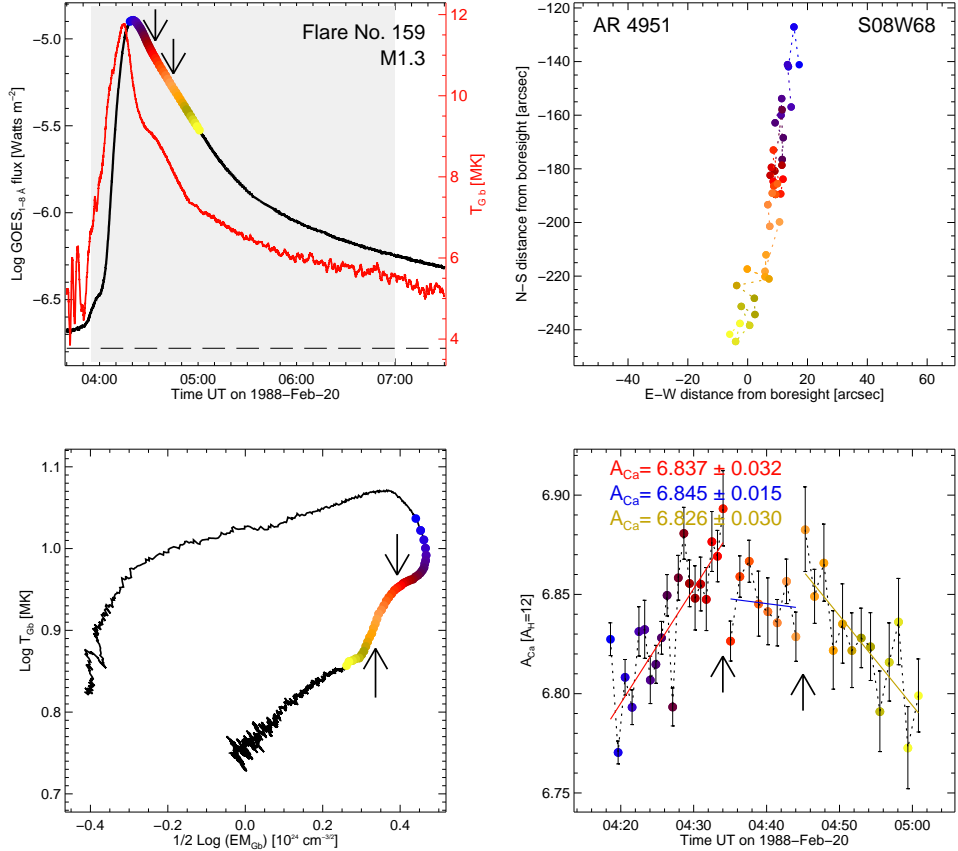}}
\caption{Four-panel diagram (like Figure~\ref{fig:fig3}) for the near-limb flare on 1988 February~20 (SOL1988-Feb-20T04:20~UT: number 159 in the on-line table and Table~\ref{tab:notable_trends}). The values of $A({\rm Ca})$ increase then decrease, with maximum values simultaneous with points of inflexion in the diagnostic diagram (bottom left panel) and a slight increase in the temperature plot (red curve, top left panel).
\label{fig:fig5}}
\end{figure}

During the M1.3 near-limb flare on 1988 February~20 (SOL1988-Feb-20T0420 UT), illustrated in Figure~\ref{fig:fig5}, $A({\rm Ca})$ initially rose steeply followed by a steady decrease in the two following segments. The changes in the time trend (04:34~UT, 04:43~UT) were simultaneous with changes in the shape of the path in the diagnostic diagram (arrowed) and slight increases in $T_{\rm Gb}$ (top left panel). The time behavior in $A({\rm Ca})$ is thus opposite to the 1984 May~19 flare, with increase followed by decrease.

%
\begin{figure}
\centerline{\includegraphics[width=0.75\textwidth,clip=]{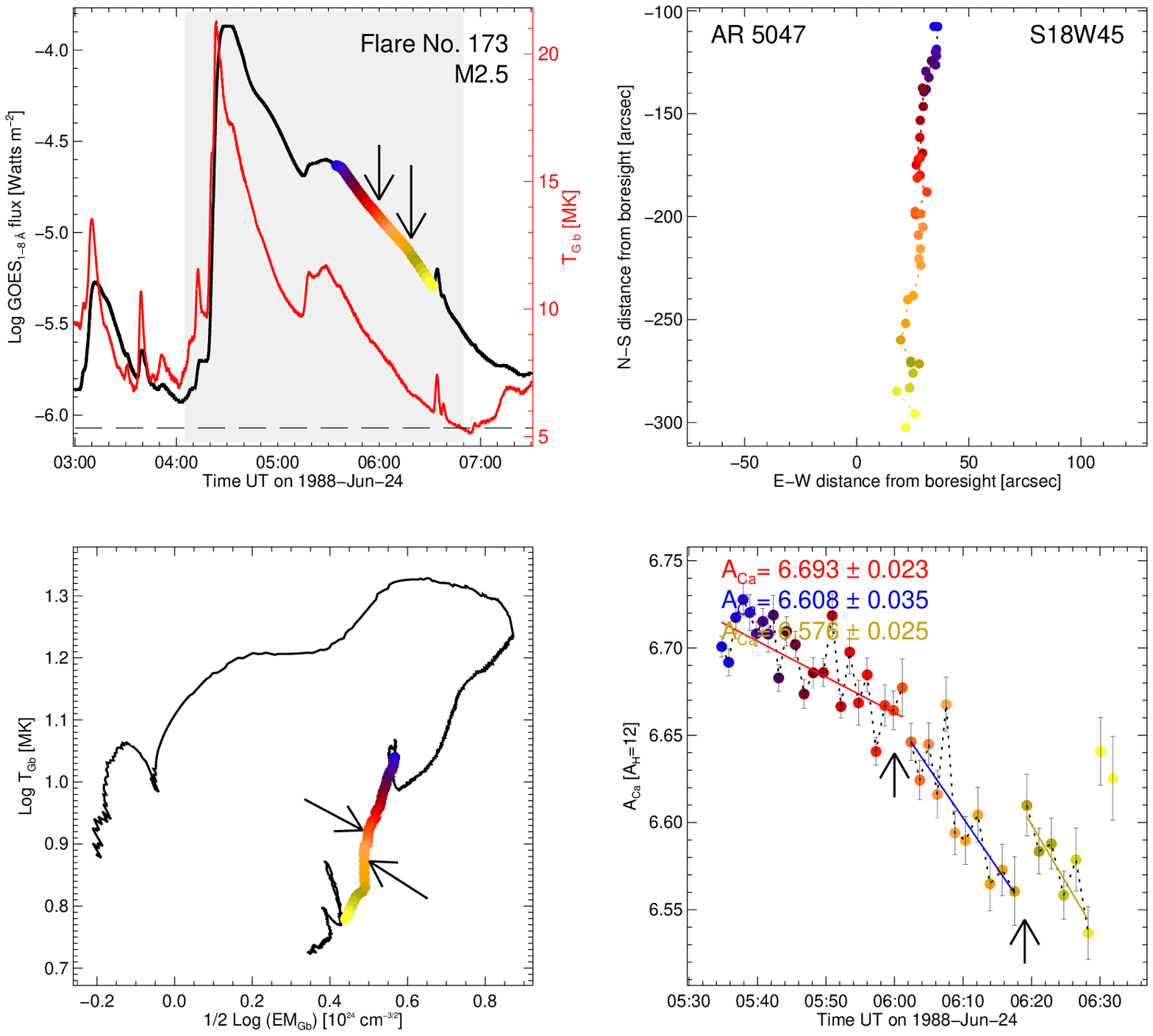}}
\caption{Four-panel diagram (like Figure~\ref{fig:fig3}) for the disk flare on 1988 June~24 (SOL1988-Jun-24T05:27~UT: number~173 in the on-line table and Table~\ref{tab:notable_trends}). For this flare, $A({\rm Ca})$ steadily decreases over a 50-minute period. The BCS measurements cover the decay of a flare starting at 05:30~UT following the main flare emission peaking at 04:30~UT. There are slight changes in the path on the diagnostic diagram indicated by arrows (bottom left panel). The main emission moves in a N~--~S direction by 3~arcminutes.
\label{fig:fig6}}
\end{figure}

BCS observations of the 1988 June~24 M2.5 disk flare (SOL1988-Jun-24T05:27 UT) were made over a 40-minute section of the long-duration flare decay (Figure~\ref{fig:fig6}). Here, the value of $A({\rm Ca})$ decreased by 0.2 in two time segments, with a large-scale motion of the flaring source (3~arcminutes) in a mostly N~--~S direction indicated. The BCS observations are over the decay of a second flare with lower temperature following a flare with peak temperature of 21~MK.

\subsection{Trends for All Flare Decays}\label{sec:Trends_for_all_flares}

\begin{figure}
\centerline{\includegraphics[width=0.75\textwidth,clip=]{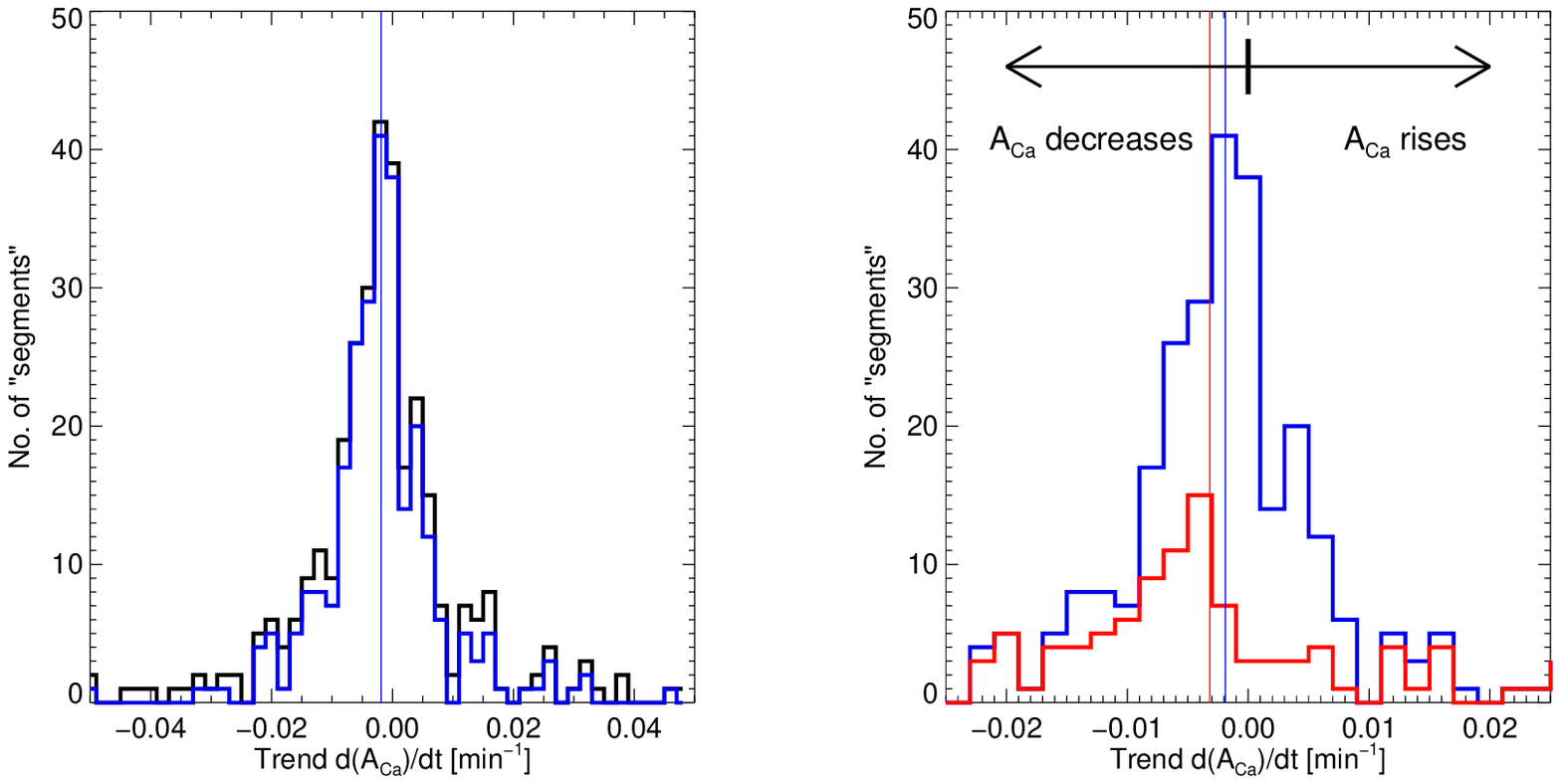}}
\caption{Left panel: Histograms of inclination trends, $d[A({\rm Ca)]}/dt$ (units of minute$^{-1}$) for all time segments analyzed (black histogram).  The distribution shown by the blue histogram includes the 268 time segments for which the estimated accuracy of the time derivative $d[A({\rm Ca)]}/dt$ is better than 2.3\%. The mean value of the trend is $ -0.0019$~minute$^{-1}$ (indicated by the thin vertical line). Right panel: Enlargement of the blue histogram (left panel) showing the best-determined time trends. The red histogram shows the distribution of the 106 time segments for which the trend is at least $3\sigma$ different from zero. The average (shown by the thin red vertical line) is $-0.0032$~minute$^{-1}$, and indicates a slight preponderance of cases when $A({\rm Ca})$ decreases with time.
\label{fig:fig7}}
\end{figure}

The distribution of the measured trends for all flare decays are shown in Figure~\ref{fig:fig7} (left panel), with the time trends segregated into two groups: those for the segments (black) identified in all flare decays and those with time derivatives $d[A({\rm Ca)]}/dt$ estimated to be precise to within 2.3\% (blue histogram). In the right panel, the time scale is enlarged and overplotted with a red histogram showing distribution of trends for flare decays having time trends of $A({\rm Ca})$ deviating by more than $3\sigma$ from zero (i.e. for which a hypothesis of constant value of $A({\rm Ca})$ abundance can be safely rejected). The average for this subclass (indicated by the red vertical line) is $-0.0032$~minute$^{-1}$. This negative value implies a slight preponderance of flare decays in which $A({\rm Ca})$ decreases with time.

%

\section{Discussion}\label{sec:Discussion}

The flare Ca abundances $A({\rm Ca})$ measured here from the emission ratios of all lines in BCS channel~1 to nearby continuum and with temperatures determined from satellite-to-resonance line ratios and GOES irradiances  have a precision that allows us to study any time variations during the course of flare decays. The four flares described in Section~\ref{sec:Example_Flares} are examples of the behavior of flare decays in which there are significant changes in $A({\rm Ca})$, with both decreases and increases with time. These flares are not meant to typify the time behaviors of flares which vary considerably in nature and are difficult to categorize. Frequently, jumps in the variations of $A({\rm Ca})$ occur, which are generally associated with discontinuities in the temperature variations or equivalently in the diagnostic diagrams.

The 270 short time segments with $A({\rm Ca})$ determined to better than 2.3\% accuracy, many (106; 39\%) showed variations in $A({\rm Ca})$ at the $3\sigma$ level.  For the majority, 74 (70\%) of these 106 segments $A({\rm Ca})$ decreased with time, and for 32 (30\%) $A({\rm Ca})$ increased with time. For 79 out of 270 (29\%) $A({\rm Ca})$ was constant or nearly constant, and the remaining 85 (31\%) with irregular time behavior.

Taking all 270 time segments of the 159 flare decays, as indicated by Figure~\ref{fig:fig7} (left panel), the time derivatives of the measured abundances, $d[A({\rm Ca)]}/dt$, have a total range of $\pm 0.04$ minute$^{-1}$. For these time segments, there is a slightly greater number showing a decreasing $A({\rm Ca})$, although the average is very small ($d[A({\rm Ca)]}/dt = -0.0019$ minute$^{-1}$). For time segments with time derivatives determined to better than $3\sigma$ the average is $d[A({\rm Ca)]}/dt = -0.0032$~minute$^{-1}$.

The theory of \cite{lam04,lam21} (see also references therein) explains the FIP effect, or enhancement of elements with first ionization potential of less than 10~eV, by a ponderomotive force associated with Alfv{\'e}n waves or magnetohydrodynamic waves traveling up or down from the photosphere. The calculations of \cite{lam21} give element abundance changes in the corona for combinations of the ratio of magnetic field strengths in the corona and photosphere ($B_{\rm cor}/B_{\rm phot}$) and fast-mode wave amplitudes $v_{fm}$ at the plasma beta $\beta = 1$ level. As stated in our previous work \citep{jsyl22}, our averaged Ca flare abundance $A({\rm Ca})$ is equal to 6.74, or about a factor of 2.6 higher than photospheric or meteoritic abundances ($A({\rm Ca}) = 6.32 \pm 0.03$). The calculations of \cite{lam21} would agree with this work for $B_{\rm cor}/B_{\rm phot} = 0.5$ and relatively small values of $v_{fm}$, less than 10~km~s$^{-1}$, which can be regarded as a ``standard'' FIP result.

If the FIP effect can be invoked to explain the results of the time-varying Ca abundances during flares as in the present work, it is improbable that parameters for the emitting plasma (plasma beta, wave velocity amplitudes) change on the short time-scales of time segments used here. It is more likely that there are changes in the flare morphology, in particular new, emerging X-ray features appearing as the main flare loops decay with their own characteristic calcium abundance $A({\rm Ca})$; each of these loops is hypothesized to have its own particular abundance with $A({\rm Ca})$ differing from the average by up to 0.2, as can be seen from Figures~\ref{fig:fig3}--\ref{fig:fig6}.

\begin{figure}
\centerline{\includegraphics[width=0.75\textwidth,clip=]{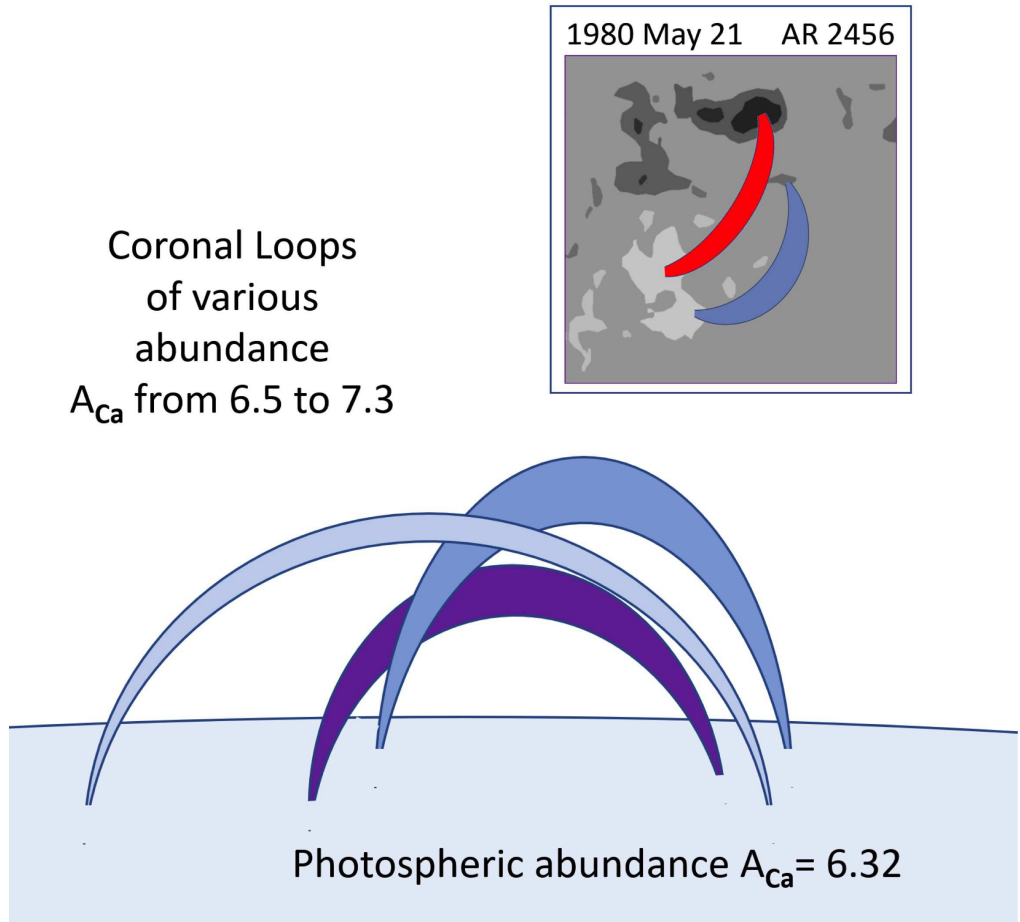}}
\caption{Interpretation of the varying calcium abundance results obtained. The dark-colored loop represents the main loop giving rise to the initial flare X-ray emission. The lighter colored loops are those that are hypothesized to emerge later in the flare development, each having its own calcium abundance which one by one dominates over the main flare loop structure. Inset: The flare loops shown schematically for the SOL1980-May-21T21:07~UT (no. 10) flare based on HXIS images for the time of the peak (red) and about an hour later (blue). The background represents a part of the Kitt Peak magnetogram for the day.
\label{fig:fig8}}
\end{figure}

A pictorial representation that we believe would explain our observations is shown in Figure~\ref{fig:fig8}. The main flare emission in the initial decay stages arises from the loop structure shown in dark-blue. As the flare evolves, other X-ray loops emerge (represented by lighter-blue colors) and dominate over the main flare emission. Each loop has its own characteristic calcium abundance $A({\rm Ca})$  which is in general different from the main loop, either smaller or larger depending on each flare. The example of the SOL1980-May-21T21:07~UT flare (no. 10) is a case in point, when the HXIS observations indicate in addition to the main loop at the time of the flare peak a loop above and to the south of the main loop about an hour later (see inset of Figure~\ref{fig:fig8} where these loops are shown schematically).  Future high-resolution X-ray images with time resolution of a minute or so, such as are being currently obtained from the Spectrometer Telescope for Imaging X-rays (STIX) instrument on Solar Orbiter, would probably help to resolve this issue.

As pointed out previously \citep{jsyl22}, the ponderomotive force model needs to take into account the apparent inverse FIP effect for Si and S which has been seen in the crystal spectrometer results of \cite{vec81} and \cite{bsyl15} and from the broad-band spectra obtained by \cite{kat20} from extremely large flares. However, time-resolved spectra from X-ray Solar Monitors on board the {\it Chandrayaan-2}, SMART-1, and MESSENGER spacecraft \citep{nar20} at present do not appear to support the earlier findings for Si and S, although it is possible that a more sophisticated differential emission model than the two-component one assumed in these works would give different results.

%

\section{Summary and Conclusions}\label{sec:Summary}

In this work we have concentrated on the special case of calcium abundances in X-ray flares since a large data set is readily available in archived form and from the BCS on SMM that is well characterized and calibrated. In addition, temperature information is available through satellite line-to-resonance line ratios in BCS spectra and the closely related temperature from the background-subtracted GOES ratios. We estimate that the precision of the Ca abundances $A({\rm Ca})$ is 0.01 for the larger flares in this study at least, and clearly the variations in Ca abundance are much larger than this. The time variations are a mixture of rising, falling, and hybrid, with frequent discontinuous jumps which appear to be related to features in the temperature--time profiles and sometimes the locations of the main flare emission as deduced from displacements of flare spectra in an E~--~W direction and the vignetting factor by comparing BCS and GOES emission for the N~--~S direction. Comparing the Ca abundances obtained here with those from the ponderomotive force theory of \cite{lam21}, a range of parameters (ratio of field strengths in the corona and photosphere, fast-mode wave amplitude) can be found that approximates our averaged measured values. It is unlikely that these parameters suddenly change with time so that the jumps in the Ca abundances seen in the flare decays we have studied are more likely due to the emergence of new loop structures as the flare evolves after its maximum. Figure~\ref{fig:fig8} represents what may occur, with loops having their own characteristic Ca abundances rising in the neighborhood of the main loop.

\begin{acknowledgments}

We acknowledge financial support from the Polish National Science Centre grant No. UMO-2017/25/B/ST9/01821. We thank the staff of the NASA Data Center Archive and Dr Dominic Zarro for reformatting XRP data. The CHIANTI atomic database and code is a collaborative project involving George Mason University, the University of Michigan (USA), and the University of Cambridge (UK).
\end{acknowledgments}

\vspace{5mm}
\facilities{Solar Maximum Mission (BCS)}
\software{SolarSoft Interactive Data Language \citep{fre98}, {\sc chianti} \citep{delz15}}

\vspace{5mm}
On-line material:

List of 194 flares from Paper 1:

{\tt http://www.cbk.pan.wroc.pl/js/ApJ\_930\_77\_2022/Table\_1.txt}

 Figures (like Figures 3 to 6) for 159 flares analysed in this study:

 {\tt http://www.cbk.pan.wroc.pl/js/ApJ\_Varying\_Coronal\_Abundances\_Files/} (see {\tt Readme.txt} there).

\bibliography{RESIK}
\bibliographystyle{aasjournal}

\end{document}